\documentclass[twocolumn]{revtex4}
\usepackage{amssymb}
\usepackage{latexsym}
\usepackage{epsfig}
\usepackage{color}
\usepackage{float}
\usepackage{graphicx}
\usepackage{epstopdf}

\begin{document}

\title{Generalized entropies and corresponding holographic dark energy models}
\author{H. Moradpour$^1$\footnote{hn.moradpour@maragheh.ac.ir}, A. H. Ziaie$^1$\footnote{ah.ziaie@maragheh.ac.ir}, M. Kord Zangeneh$^2$\footnote{mkzangeneh@scu.ac.ir}}
\address{$^1$ Research Institute for Astronomy and Astrophysics of Maragha (RIAAM), University of Maragheh, P.O. Box 55136-553, Maragheh, Iran\\
$^2$ Physics Department, Faculty of Science, Shahid Chamran
University of Ahvaz, Ahvaz 61357-43135, Iran}

\begin{abstract}
Using Tsallis statistics and its relation with Boltzmann entropy, the Tsallis entropy content of black holes is achieved, a result in full agreement with a recent study (Phys. Lett. B 794, 24 (2019)). In addition, employing Kaniadakis statistics and its relation with that of Tsallis, the Kaniadakis entropy of black holes is obtained. The Sharma-Mittal and R\'{e}nyi entropy contents of black holes are also addressed by employing their relations with Tsallis entropy. Thereinafter, relying on the holographic dark energy hypothesis and the obtained entropies, two new holographic dark energy models are introduced
and their implications on the dynamics of a flat FRW universe are studied when there is also a pressureless fluid in background. In our setup, the apparent horizon is considered as the IR cutoff, and there is not any mutual interaction between the cosmic fluids. The results indicate that the obtained cosmological models have $i$) notable powers to describe the cosmic evolution from the matter-dominated era to the current accelerating universe, and $ii$) suitable predictions for the universe age.
\end{abstract}

\maketitle

\section{Introduction}

Originally, Gibbs put forth that systems including long-range
interactions may not be extensive \cite{gibbs}, a hypothesis which also
motivates people to propose various entropy definitions
\cite{Kani,pla}. Recently, such entropies have been employed to
model the cosmic evolution in various setups
\cite{non13,nonK,EPJC,non20,eplT}. Generalized entropies have also
been employed to build new holographic dark energy models
\cite{Tavayef,epjcr,smm}. It has been shown that such entropies can
also provide a theoretical basis for the MOND theory \cite{plbmond},
affect the Jeans mass \cite{mnras}, may be motivated by the quantum
features of gravity \cite{eplm}, and even describe inflation without
considering inflaton \cite{rinf}.

Bekenstein entropy \cite{ser} (and therefore, the nature of degrees
of freedom of horizon \cite{pav,pav1}) is the cornerstone of
Holographic dark energy hypothesis (HDE) \cite{HDE}, a promising
approach to understand the origin of dark energy. Motivated by
recent works on cosmological and gravitational consequences of
Kaniadakis statistics \cite{nonK,eplK}, as a generalized entropy
measure \cite{Kani,pla}, and various features and properties of
previously introduced generalized entropy based holographic dark energy models
\cite{Tavayef,epjcr,smm,clas}, here, we are going to $i$) calculate the entropy of black holes in the various well-known generalized entropy formalisms, and $ii$) study their ability in describing the current accelerated universe by building their corresponding holographic dark energy models in the unit of $k_B=G=c=\hbar=1$.

In the next section, after addressing relation between Tsallis and Boltzmann entropies, Tsallis entropy of black holes is derived, a result compatible with that of Ref.~\cite{mah2019}. In addition, having introduced Kaniadakis entropy and its relation with Tsallis entropy, we also
compute the Kaniadakis entropy of black hole in the next section. Focusing on the relations between Tsallis entropy and Sharma-Mittal and R\'{e}nyi entropies \cite{smm,epjcr}, the Sharma-Mittal and R\'{e}nyi entropy contents of black holes are other issues which will also be addressed in the second section.
The Kaniadakis holographic dark energy (KHDE) shall be introduced in
Sec.~(\textmd{III}), where some of its cosmological consequences are
also investigated. The ability of obtained Tsallis entropy in describing dark energy as the vacuum energy, through constructing the corresponding HDE model, is also studied in Sec.~($\textmd{IV}$). The universe age is finally addressed in the fifth section. In the last section, some concluding remarks have
been collected.
\section{Tsallis and Kaniadakis entropies of black holes}

Working in the unit $k_B=1$, both the Shannon and Gibbs entropies of a
distribution with $W$ states lead to the same expression

\begin{eqnarray}\label{gib}
&&S=-\sum_{i=1}^{W}P_i\ln(P_i),
\end{eqnarray}

\noindent while $P_i$ denotes the probability
of occupying the $i$th state, for the classical systems. The quantum mechanical version of
this entropy, the so-called Von-Neumann entropy, is also presented
as

\begin{eqnarray}\label{von}
&&S=-\texttt{T}r\big[\rho\ln(\rho)\big].
\end{eqnarray}

\noindent The use of Eq.~(\ref{von})
for classical systems goes indeed back to the Boltzmann's proposal where $\rho$ is
the state density in phase space \cite{bolt}.

Applying Eq.~(\ref{von}) to a purely gravitational system, the so-called Bekenstein entropy $S_{BH}=\frac{A}{4}$ is obtainable \cite{ser}. Since the degrees of
freedom are distributed on horizon without any specific priority
with respect to one another, one may assume that, at least based on
our knowledge \cite{pav,pav1}, $P_i$ is equal for all of them allowing
us in writing $P_i=\frac{1}{W}$. In this manner, both of the above
relations lead to the Boltzmann entropy ($S=\ln(W)$), and hence, we
have \cite{eplK,eplT}

\begin{eqnarray}\label{SS}
S_{BH}=\frac{A}{4}=\ln(W)\rightarrow W=\exp(\frac{A}{4}),
\end{eqnarray}

\noindent for the horizon entropy, and consequently $W(A)$.

The Tsallis entropy, as a single-free parameter generalized entropy,
is defined as \cite{pla}

\begin{eqnarray}\label{Te}
&&S_Q^T=\frac{1}{1-Q}\sum_{i=1}^{W}(P_i^Q-P_i)=\frac{W^{1-Q}-1}{1-Q},
\end{eqnarray}

\noindent where $Q$, named Tsallis or non-extensive parameter, is an unknown free parameter ($S_Q^T\rightarrow
S$ for $Q\rightarrow1$), and the last line is valid only for
probability distributions meeting the $P_i=\frac{1}{W}$ condition.
The $Q$ parameter may also be arisen from the quantum features of
gravity \cite{eplK,eplm}. Now, using Eq.~(\ref{SS}) and the last line of Eq.~(\ref{Te}), one can
easily find

\begin{eqnarray}\label{ss1}
&&S_Q^T=\frac{1}{1-Q}[\exp\big((1-Q)S_{BE}\big)-1].
\end{eqnarray}

In the loop quantum gravity scenario, by
applying Tsallis entropy definition to black holes, it is obtained that
\cite{mah2019}

\begin{eqnarray}\label{Te1}
&&S_Q^T=\frac{1}{1-Q}[\exp\big(\frac{(1-Q)\ln(2)}{\gamma\pi\sqrt{3}}S_{BE}\big)-1],
\end{eqnarray}

\noindent satisfying the $S_Q^T\rightarrow S_{BE}$ expectation
whenever $Q\rightarrow1$ and $\gamma=\frac{\ln(2)}{\pi\sqrt{3}}$
\cite{mah2019}. Therefore, in order to preserve the
$S_Q^T\rightarrow S_{BE}$ expectation at the $Q\rightarrow1$
limit, we consider $\gamma=\frac{\ln(2)}{\pi\sqrt{3}}$, and accordingly, Eqs.~(\ref{Te1}) and~(\ref{ss1}) become the same.

The Kaniadakis entropy ($\kappa$-entropy), as another single-free parameter
generalized entropy, is also defined as \cite{Kani}

\begin{eqnarray}\label{Ke}
&&S_\kappa=-\sum_{i=1}^{W}\frac{P_i^{1+\kappa}-P_i^{1-\kappa}}{2\kappa}=\\
&&\frac{1}{2}\bigg(\frac{\sum_{i=1}^{W}(P_i^{1-\kappa}-P_i)}{\kappa}+\frac{\sum_{i=1}^{W}(P_i^{1+\kappa}-P_i)}{-\kappa}\bigg),\nonumber
\end{eqnarray}

\noindent where $\kappa$ is an unknown parameter, and the Boltzmann-Gibbs entropy is recovered at the $\kappa\rightarrow0$ limit \cite{Kani}. Comparing Eqs.~(\ref{Ke}) and~(\ref{Te}) with each other, one can easily obtain

\begin{eqnarray}\label{Ke1}
&&S_\kappa=\frac{S_{1+\kappa}^T+S_{1-\kappa}^T}{2}.
\end{eqnarray}

\noindent Moreover, by assuming $P_i=\frac{1}{W}$, Eq.~(\ref{Ke})
helps us in getting \cite{Kani}

\begin{eqnarray}\label{Ke0}
&&S_\kappa=\frac{W^{\kappa}-W^{-\kappa}}{2\kappa},
\end{eqnarray}

\noindent combined with Eq.~(\ref{SS}) to reach at

\begin{eqnarray}\label{Ke01}
&&S_\kappa=\frac{1}{\kappa}\sinh\big(\kappa S_{BE}).
\end{eqnarray}

\noindent Indeed, putting Eq.~(\ref{Te1}) or Eq.~(\ref{ss1}) in
Eq.~(\ref{Ke1}), one can again reach this result, the Kaniadakis
entropy of a black hole.

Now, as a brief study, let us investigate the Sharma-Mittal and R\'{e}nyi entropy contents of a black hole with Tsallis entropy~(\ref{ss1}). Since Sharma-Mittal ($S_{SM}$) and R\'{e}nyi ($\mathcal{S}$) entropies can be expressed as the functions of Tsallis entropy as \cite{smm,epjcr}

\begin{eqnarray}\label{reyn1}
&&S_{SM}=\frac{1}{R}\big((1+(1-Q)S_T)^{\frac{R}{1-Q}}-1\big),\nonumber\\
&&\mathcal{S}=\frac{1}{1-Q}\ln(1+(1-Q)S_T),
\end{eqnarray}

\noindent respectively, in which $R$ is an unknown parameter \cite{smm,epjcr}, a straightforward calculation leads to

\begin{eqnarray}
&&S_{SM}=\frac{1}{R}[\exp\big(RS_{BE}\big)-1],\nonumber\\
&&\mathcal{S}=S_{BE},
\end{eqnarray}

\noindent where we assumed $\gamma=\frac{\ln(2)}{\pi\sqrt{3}}$, otherwise $\frac{\ln(2)}{\gamma\pi\sqrt{3}}S_{BE}$ would emerge in results instead of $S_{BE}$. The obtained $S_{SM}$ and $\mathcal{S}$ are mathematically equal to Tsallis entropy~(\ref{ss1}) and Bekenstein entropy, respectively, meaning that they can not tell us anything more than what Tsallis entropy~(\ref{ss1}) and Bekenstein entropy give us in constructing HDE models. Hence, we only focus on Eqs.~(\ref{Ke01}) and~(\ref{ss1}) to address two new HDE models.

\section{KHDE}

Bearing the Kaniadakis entropy content of a black hole in
mind~(\ref{Ke01}), since the HDE hypothesis claims that if vacuum
energy handles the current accelerated universe, then its amount
stored in a box with size $L^3$ should not exceed the energy of its
same size black hole \cite{HDE}, one can reach

\begin{eqnarray}\label{coh}
\Lambda^4\equiv\rho^\kappa_\Lambda\propto\frac{S_{\kappa}}{L^4},
\end{eqnarray}

\noindent for the vacuum energy $\rho^\kappa_\Lambda$, finally leading to

\begin{eqnarray}\label{energydensity}
\rho^\kappa_\Lambda=\frac{3C^2H^4}{8\pi\kappa}\sinh(\frac{\pi\kappa}{H^2}),
\end{eqnarray}

\noindent where $C^2$ is an unknown constant as usual, and the apparent horizon of flat FRW universe is considered
as the IR cutoff \cite{HDE}. Now, it is apparent that whenever $\kappa\rightarrow0$, we have $\rho^\kappa_\Lambda\rightarrow\frac{3C^2H^2}{8\pi}$, the well-known Bekenstein entropy-based HDE \cite{HDE}. In the absence of a mutual interaction
between the cosmic sectors, including a pressureless fluid with
energy density $\rho_m$ and the dark energy candidate with energy
density $\rho^\kappa_\Lambda$ and pressure $p_\Lambda$, the energy-momentum
conservation law and Friedmann equations take the forms

\begin{figure}[ht]
\begin{center}
\includegraphics[scale=0.47]{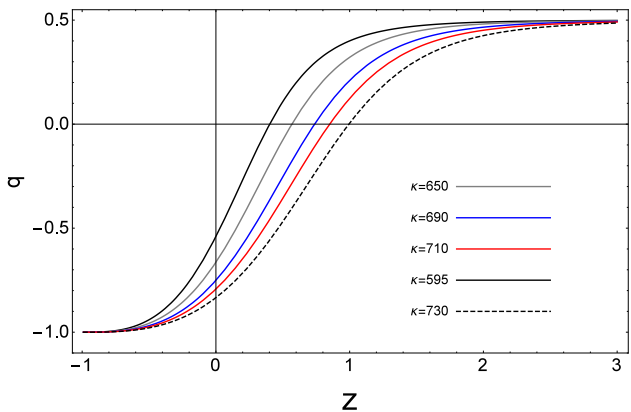}
\caption{The behavior of deceleration parameter for $C^2=0.3025$,
$\Omega_m^0=0.315$ (the current value of $\Omega_m$), and
$H_0=67.9$ \cite{planck}.}\label{fig1}
\end{center}
\end{figure}

\begin{eqnarray}\label{emc1}
&&\dot{\rho}_{m}+3H\rho_{m}=0,\\
&&p^\kappa_\Lambda=-(\frac{\dot{\rho}^\kappa_\Lambda}{3H}+\rho^\kappa_\Lambda),\nonumber
\end{eqnarray}

\noindent and

\begin{eqnarray}\label{fe1}
&&H^2=\frac{8\pi}{3}(\rho_m+\rho^\kappa_\Lambda),\\
&&H^2+\frac{2}{3}\dot{H}=\frac{-8\pi}{3}(p^\kappa_\Lambda),\nonumber
\end{eqnarray}

\noindent respectively, where dot denotes derivative with respect to
time. Deceleration parameter $q$ is also obtainable as

\begin{eqnarray}\label{dec1}
q=-1-\frac{\dot{H}}{H^{2}}=\frac{1}{2}(1+\frac{3\omega^\kappa_\Lambda}{\frac{\Omega_{m}}{\Omega^\kappa_\Lambda}+1}),
\end{eqnarray}

\noindent in which

\begin{eqnarray}\label{3}
&&\Omega^\kappa_\Lambda\equiv\frac{\rho^\kappa_\Lambda}{\rho_c}=\frac{C^2H^2}{\kappa}\sinh(\frac{\pi\kappa}{H^2}),\\
&&\Omega_{m}\equiv\frac{\rho_{m}}{\rho_c},\
\omega^\kappa_\Lambda\equiv\frac{p^\kappa_\Lambda}{\rho^\kappa_\Lambda},\ \rho_c=\frac{3H^2}{8\pi}.\nonumber
\end{eqnarray}

Evolution of $q$ versus redshift $z$ has been plotted in
Fig.~(\ref{fig1}), which clearly shows that, depending on the values
of $C^2$ and $\kappa$, a desired behavior is getable. These values
can also affect the value of transition redshift $z_{tr}$ at which
$q=0$ and universe changes its acceleration phase.

\begin{figure}[ht]
\begin{center}
\includegraphics[scale=0.35]{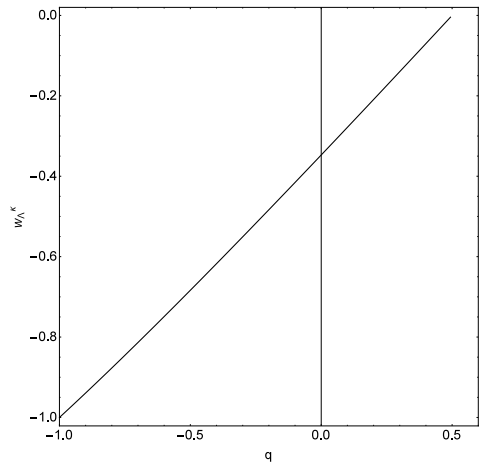}
\caption{The behavior of $w^\kappa_\Lambda$ against deceleration parameter
for $C^2=0.3025$, $\Omega_m^0=0.315$, $H_0=67.9$ and
$\kappa=690$.}\label{fig3}
\end{center}
\end{figure}

$w^\kappa_\Lambda$ is also depicted as a function of $q$ only for
$\kappa=690$ in Fig.~(\ref{fig3}), because the curves
corresponding to other values of $\kappa$, used in plotting
Fig.~(\ref{fig1}), are so close to this curve. As it is apparent,
KHDE behaves as a pressureless fluid for high redshift limits (when
$q=\frac{1}{2}$) and mimics a cosmological constant ($\equiv
w^\kappa_\Lambda\rightarrow-1$) when $z\rightarrow-1$ for which $q\rightarrow-1$ and $\rho_m\rightarrow0$.

\begin{figure}[ht]
\begin{center}
\includegraphics[scale=0.45]{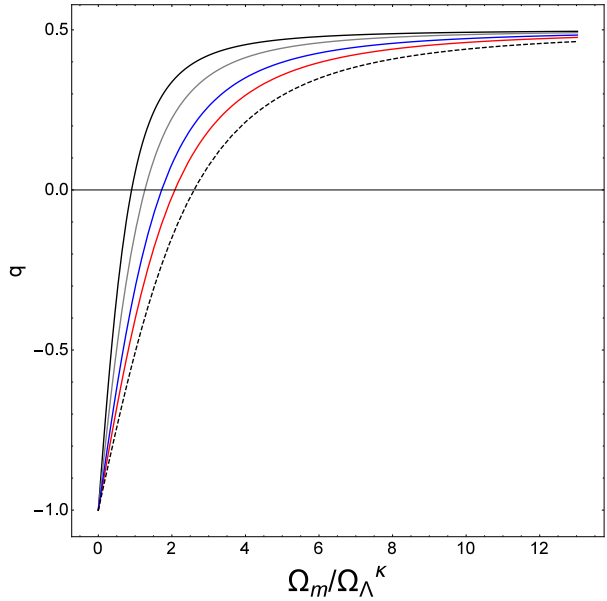}
\caption{The behavior of deceleration parameter against the ratio
$\Omega_m/\Omega^\kappa_\Lambda$ for $C^2=0.3025$, $\Omega_m^0=0.315$,
$H_0=67.9$. We have $\Omega_m/\Omega^\kappa_\Lambda\approx0.95$ at
$z_{tr}\approx0.570$ for gray curve and
$\Omega_m/\Omega^\kappa_\Lambda\approx2.01$ at $z_{tr}\approx0.997$ for
black dashed curve.}\label{fig2}
\end{center}
\end{figure}

In fact, by decreasing $z$, KHDE starts to control the universe
expansion rate and approaches its maximum ($1$) when
$z\rightarrow-1$ or equally $q\rightarrow-1$. At high redshift
limits, the energy density of matter density begins to become
dominant and matter dominated era is begun to be covered i.e. we have
$q\rightarrow\frac{1}{2}$ and
$\frac{\Omega^\kappa_\Lambda}{\Omega_m}\rightarrow0$. In order to have a
better look, we plotted $q(\frac{\Omega_m}{\Omega^\kappa_\Lambda})$ in
Fig.~(\ref{fig2}). As it is apparent, the ratio $\frac{\Omega_m}{\Omega^\kappa_\Lambda}$ increases as a function of $z$, and the matter dominated era ($q=\frac{1}{2}$) shall be recovered at high redshift limit in full agreement with what one observes in previous figures. At the transition point, we have $q=0$ leading to

\begin{eqnarray}
\frac{\Omega_m}{\Omega^\kappa_\Lambda}=-1-3\omega^\kappa_\Lambda,
\end{eqnarray}

\noindent which states that $\frac{\Omega_m}{\Omega^\kappa_\Lambda}>0$ only
if $\omega_\Lambda<-\frac{1}{3}$. For example, for
$\Lambda\textmd{CDM}$ model, we have
$\frac{\Omega_m}{\Omega^\kappa_\Lambda}=2$. In Fig.~(\ref{fig2}), a variety of
the $\frac{\Omega_m}{\Omega^\kappa_\Lambda}$ ratio is obtainable, depending on
the values of $C^2$ and $\kappa$ parameters.

\section{New Tsallis Holographic dark energy}

Following the above approach, and by using Tsallis entropy~(\ref{ss1}) instead of Kaniadakis entropy, one reaches a new Tsallis Holographic dark energy (NTHDE) expressed as

\begin{eqnarray}\label{en1}
\rho^T_\Lambda=(\frac{3D^2}{8\pi})\frac{2H^4}{\delta\pi}\exp(\frac{\delta\pi}{2H^2})\sinh(\frac{\delta\pi}{2H^2}),
\end{eqnarray}
\begin{figure}[ht]
	\begin{center}
		\includegraphics[scale=0.45]{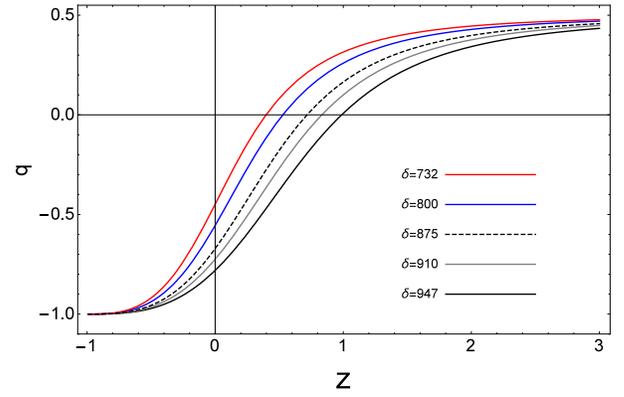}
		\caption{The behavior of deceleration parameter for $D^2=0.3136$,
			$\Omega_m^0=0.315$ (the current value of $\Omega_m$), and
			$H_0=67.9$.}\label{fig4}
	\end{center}
\end{figure}

\noindent in which $D^2$ is an unknown constant as usual, and $\delta=1-Q$. In this manner, Eqs.~(\ref{emc1})-(\ref{3}) are still valid if we apply the below changes to them
\begin{eqnarray}
&&\rho^\kappa_\Lambda\Rightarrow\rho^T_\Lambda,\ p^\kappa_\Lambda\Rightarrow p^T_\Lambda,\
\omega^\kappa_\Lambda\Rightarrow\omega^T_\Lambda\equiv\frac{p^T_\Lambda}{\rho^T_\Lambda},\\
&&\Omega^\kappa_\Lambda\Rightarrow\Omega^T_\Lambda\equiv\frac{\rho^T_\Lambda}{\rho_c}=\frac{2D^2}{\delta\pi}H^2\exp(\frac{\delta\pi}{2H^2})\sinh(\frac{\delta\pi}{2H^2}).\nonumber
\end{eqnarray}

\begin{figure}[ht]
	\begin{center}
		\includegraphics[scale=0.4]{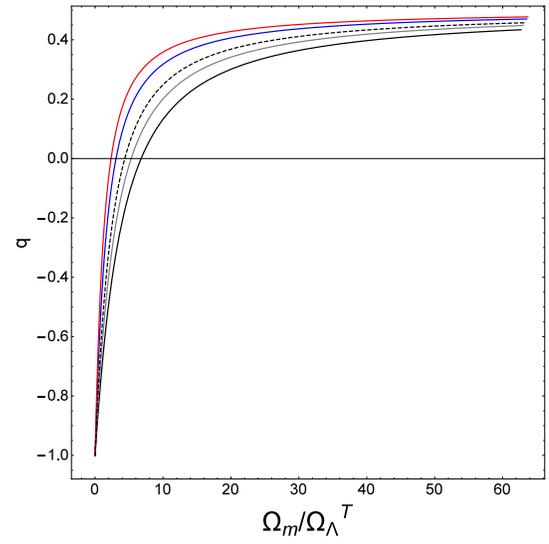}
		\caption{The behavior of deceleration parameter against the ratio
			$\Omega_m/\Omega^T_\Lambda$ for $D^2=0.3136$, $\Omega_m^0=0.315$, and
			$H_0=67.9$.}\label{fig5}
	\end{center}
\end{figure}

In Figs.~(\ref{fig4})-(\ref{fig5}), deceleration parameter is plotted against redshift and ratio $\frac{\Omega_m}{\Omega^T_\Lambda}$, respectively, indicating that this ratio increases as a function of $z$, and matter dominated era is achievable by increasing redshift.
The same as KHDE, NTHDE behaves as pressureless source when $q=\frac{1}{2}$, and mimics cosmological constant ($w^T_\Lambda=-1$) for $q\rightarrow-1$. We also plotted $w^T_\Lambda$ against $q$ only for $\delta=947$, because of the proximity of the curves corresponding on other values of $\delta$ to each other, a property also seen in the case of KHDE.
\begin{figure}[ht]
	\begin{center}
		\includegraphics[scale=0.35]{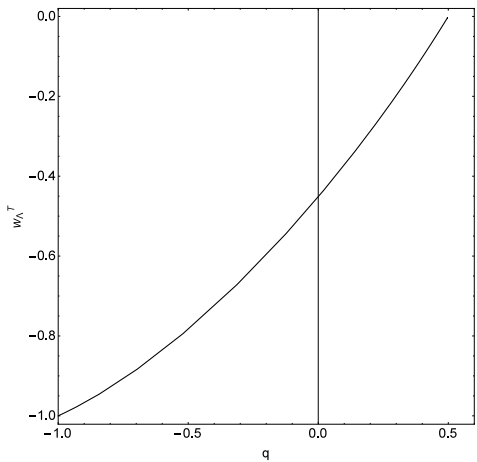}
		\caption{The behavior of $w^T_\Lambda$ against deceleration parameter
			for $D^2=0.3136$, $\Omega_m^0=0.315$, $H_0=67.9$ and
			$\delta=947$.}\label{fig6}
	\end{center}
\end{figure}

\section{The Universe age}

In order to find the predictions of the above models for the universe age, one can write

\begin{eqnarray}\label{0}
&&t=\int\frac{dtdH}{dH}=\int\frac{dH}{\dot{H}}=-\frac{1}{4\pi}\int\frac{dH}{\frac{3H^2}{8\pi}+p_\Lambda},
\end{eqnarray}

\noindent where $p_\Lambda$ denotes the pressure of the introduced HDE models, i.e. KHDE and NTHDE. As it is apparent from the behavior of KHDE and NTHDE, we have $0\leq-p_\Lambda<\rho_\Lambda<\rho_c$ when $-1<z$. In this regard, one may estimate Eq.~(\ref{0}) as

\begin{eqnarray}\label{01}
\!\!\!\!\! t_{ub}\approx\frac{-2}{3(1+w_\Lambda\Omega_\Lambda)_{z=0}}\int\frac{dH}{H^2}=\frac{2}{3H_0}(1+w_\Lambda\Omega_\Lambda)^{-1}_{z=0}.
\end{eqnarray}

\noindent Indeed, for the models, the corresponding pressure approaches zero at high redshift limit, and takes negative values by decreasing $z$. It means that the $\frac{1}{\frac{3H^2}{8\pi}+p_\Lambda}$ term always increases during the cosmic evolution. Therefore, in order to reach the above primary estimation, we replaced this term with its current value (i.e. $(1+w_\Lambda\Omega_\Lambda)_{z=0}$), and pull it outside the integral. By doing so, an upper bound, depending on the values of model parameters, for the age of current universe ($t_{ub}$) can be achieved. The same approximation has also been used in order to guess the universe age in other models \cite{Tavayef,clas}.

Now, let us consider KHDE, for which, depending on the values of $\kappa$ used to plot the curves, we have $-0.925<w^\kappa_\Lambda(z=0)<-0.74$ when $C^2=0.3025$. In this manner, one finds that the upper bound of the universe age is in the range of $(\frac{2}{3H_0}<)\frac{1.35}{H_0}<t^\kappa_{ub}<\frac{1.82}{H_0}$. For the second case, one reaches $-0.93<w^T_\Lambda(z=0)<-0.76$, and hence, $\frac{1.39}{H_0}<t^T_{ub}<\frac{1.84}{H_0}$ as the allowed range for the upper bound of the universe age when $\delta$ lies within the range employed to plot the corresponding curves in previous section, and $D^2=0.3136$. Since we used $\Omega_m^0=0.315$ in order to plot the diagrams in our investigation, we have $\Omega_\Lambda^0=1-\Omega_m^0=0.685$ for the current value of density parameter of dark energy candidate in the models. In both models, the obtained upper bound of the universe age also increases as the current value of state parameter decreases.

On the other hand, during the cosmic evolution, $p_\Lambda\leq0$ and its maximum value (zero) is achieved at high redshift limit. Therefore, at each point $H$, we have $\frac{1}{\frac{3H^2}{8\pi}}\leq\frac{1}{\frac{3H^2}{8\pi}+p_\Lambda}$ meaning that if one estimates $\frac{1}{\frac{3H^2}{8\pi}+p_\Lambda}$ with $\frac{1}{\frac{3H^2}{8\pi}}$, then Eq.~(\ref{0}) gives

\begin{eqnarray}
t_{lb}\approx-\frac{2}{3}\int\frac{dH}{H^2}=\frac{2}{3H_0}<-\frac{1}{4\pi}\int\frac{dH}{\frac{3H^2}{8\pi}+p_\Lambda},
\end{eqnarray}

\noindent indeed, a lower bound on the universe age in the models. In summary, the models predictions about the universe age lies within the range $t_{lb}<t<t_{ub}$. Although the lower bound $t_{lb}$ is common in the models, the value of $t_{ub}$ for each case is different, it depends on the values of the models parameters, and for the parameters values addressed in the previous sections, it is greater than the Hubble time $t_H$ (i.e. $t_H=\frac{1}{H_0}<t_{ub}$).

It seems that although the age of HD$140283$ ($t_{HD}$) is very close to the Hubble time, due to the accuracy of current measurements, it is still defensible in the framework of the Planck data which predicts a universe younger than $\frac{1}{H_0}$ \cite{star,planck}. In fact, the Planck estimation of the universe age ($t_p$) is something about $\frac{0.95}{H_0}$ \cite{planck}. Therefore, $t_{lb}<t_p\sim t_H\approx t_{HD}<t_{ub}$ meaning that, depending on the values of the models parameters, the models can be free of the age problem.

\section{Concluding Remarks}

Relying on relation between Tsallis and Boltzmann entropies, and assuming that all degrees of freedom of horizon have same probability, we could obtain the Tsallis entropy of black holes, a result fully compatible with that of Ref.~\cite{mah2019}. Thereafter, using the Kaniadakis statistics and its relation with Tsallis statistics, we got the Kaniadakis entropy content of black holes. The Sharma-Mittal and R\'{e}nyi entropy contents of black holes have also been calculated, and we saw that R\'{e}nyi entropy leads to Bekenstein entropy. On the other hand, although Sharma-Mittal entropy is a two free parameters entropy measure \cite{pla}, we found out that its final estimation of black hole entropy is mathematically similar to that of Tsallis entropy.

Next, applying the HDE hypothesis to the obtained Kaniadakis entropy, a new HDE model (KHDE) is derived. Our study shows that KHDE can model the current accelerated universe, and furthermore,
suitable transition from the matter dominated era to the current era is achievable by choosing proper values for the model parameters including $C^2$ and
$\kappa$. We also used the Tsallis entropy~(\ref{ss1}) to construct a new Tsallis HDE model, in short NTHDE, and studied its cosmological outcomes. It has been found out that its general behavior is relatively like that of KHDE, and depending on the values of $\delta$ and $D^2$, different transition redshifts are achievable.

Finally, we addressed the models estimations of allowed ranges for the universe age, and found out that the age problem can be avoided in these models, depending on the values of the models parameters including ($\kappa$, $C^2$) for KHDE, and ($\delta$, $D^2$) for the second one. The validity of $\frac{1}{\frac{3H^2}{8\pi}}\leq\frac{1}{\frac{3H^2}{8\pi}+p_\Lambda}=\frac{(1+w_\Lambda\Omega_\Lambda)^{-1}}{\frac{3H^2}{8\pi}}\leq\frac{(1+w_\Lambda\Omega_\Lambda)_{z=0}^{-1}}{\frac{3H^2}{8\pi}}$, due to the facts that $i$) $\dot{H},p_\Lambda<0$, and $ii$) $-p_\Lambda<\rho_\Lambda<\rho_c$ during the cosmic evolution, was the backbone of our estimations of the universe age.


\end{document}